\documentclass[a4paper,11pt,titlepage,oneside]{book} 
\pdfoutput=1

\usepackage[T1]{fontenc}
\usepackage[utf8]{inputenc}

\usepackage{ae}               
\usepackage[pdftex]{graphicx}
\usepackage{vmargin}          
\usepackage{fancyhdr}        
\usepackage[font=footnotesize]{subfig}
\usepackage[hyphens]{url}
\usepackage[absolute,overlay]{textpos}
\usepackage{tikz}
\usepackage[english]{babel}
\usepackage{color}
\usepackage{longtable}
\usepackage{cite}
\usepackage[bottom]{footmisc}
\usepackage[section,subsection,subsubsection]{extraplaceins}
\usepackage{multirow}
\usepackage{array}
\usepackage{nomencl}
\usepackage{color}
\usepackage{xcolor}
\usepackage{setspace}
\usepackage{listings}
\usepackage{microtype}
\usepackage{paralist}
\usepackage{color}
\definecolor{light-gray}{gray}{0.98}

\makeatletter
\newcommand\footnoteref[1]{\protected@xdef\@thefnmark{\ref{#1}}\@footnotemark}
\makeatother

\makenomenclature

\makeglossary

\definecolor{darkred}{rgb}{0.5,0,0}
\definecolor{darkgreen}{rgb}{0,0.5,0}
\definecolor{darkblue}{rgb}{0,0,0.5}

\setmarginsrb
{2.5cm}   
{1.5cm}   
{2.5cm}  
{2cm}   
{20pt}    
{0.25in}  
{9pt}     
{0.3in}   

\usepackage{chngcntr}
\counterwithin{figure}{section}
\counterwithin{table}{section}

\usepackage{authblk}

\newcommand\mytitle{Technical Information on Vulnerabilities of Hypercall Handlers}
\title{Cloud Usage Patterns}

\newcommand\TRnumber{Technical Report: SPEC-RG-2014-001\\Version: 1.0}
\newcommand\WGname{SPEC RG IDS Benchmarking Working Group}

\newcommand\TRdate{August 23, 2014}
\newcommand\TRcentralURL{research.spec.org}
\newcommand\TRrightURL{www.spec.org}

\newcommand\numAuthors{5} 
\makeatletter
\newcommand\defcase[1]{\@namedef{mycase@\the\numexpr#1\relax}}
\newcommand\putAuthors[1]{\@nameuse{mycase@\the\numexpr#1\relax}}
\makeatother

\defcase{0}{
	\begin{textblock}{8}[0,0](\colsinglecentralX,\rowoneY)
		\centering
		\small{\textbf{No Author}}\\
	\end{textblock}	
}

\defcase{1}{
	\begin{textblock}{8}[0,0](\colsinglecentralX,\rowoneY)
		\centering
		\small{\textbf{\authorOneName}}\\
		\footnotesize
		\authorOneAffil
	\end{textblock}	
}
\defcase{2}{
	\begin{textblock}{\authorCellWidth}[0,0](\colDoubleLeftX,\rowoneY)
		\centering
		\small{\textbf{\authorOneName}}\\
		\footnotesize
		\authorOneAffil
	\end{textblock}	
	\begin{textblock}{\authorCellWidth}[0,0](\colDoubleRightX,\rowoneY)
		\centering
		\small{\textbf{\authorTwoName}}\\
		\footnotesize
		\authorTwoAffil
	\end{textblock}	
}
\defcase{3}{
	\begin{textblock}{\authorCellWidth}[0,0](\coloneX,\rowoneY)
		\centering
		\small{\textbf{\authorOneName}}\\
		\footnotesize
		\authorOneAffil
	\end{textblock}	
	\begin{textblock}{\authorCellWidth}[0,0](\coltwoX,\rowoneY)
		\centering
		\small{\textbf{\authorTwoName}}\\
		\footnotesize
		\authorTwoAffil
	\end{textblock}	
	\begin{textblock}{\authorCellWidth}[0,0](\colthreeX,\rowoneY)
		\centering
		\small{\textbf{\authorThreeName}}\\
		\footnotesize
		\authorThreeAffil
	\end{textblock}	
}
\defcase{4}{
	\begin{textblock}{\authorCellWidth}[0,0](\coloneX,\rowoneY)
		\centering
		\small{\textbf{\authorOneName}}\\
		\footnotesize
		\authorOneAffil
	\end{textblock}	
	\begin{textblock}{\authorCellWidth}[0,0](\coltwoX,\rowoneY)
		\centering
		\small{\textbf{\authorTwoName}}\\
		\footnotesize
		\authorTwoAffil
	\end{textblock}	
	\begin{textblock}{\authorCellWidth}[0,0](\colthreeX,\rowoneY)
		\centering
		\small{\textbf{\authorThreeName}}\\
		\footnotesize
		\authorThreeAffil
	\end{textblock}
	\begin{textblock}{8}[0,0](\colsinglecentralX,\rowtwoY)
		\centering
		\small{\textbf{\authorFourName}}\\
		\footnotesize
		\authorFourAffil
	\end{textblock}	
}
\defcase{5}{
	\begin{textblock}{\authorCellWidth}[0,0](\coloneX,\rowoneY)
		\centering
		\small{\textbf{\authorOneName}}\\
		\footnotesize
		\authorOneAffil
	\end{textblock}	
	\begin{textblock}{\authorCellWidth}[0,0](\coltwoX,\rowoneY)
		\centering
		\small{\textbf{\authorTwoName}}\\
		\footnotesize
		\authorTwoAffil
	\end{textblock}	
	\begin{textblock}{\authorCellWidth}[0,0](\colthreeX,\rowoneY)
		\centering
		\small{\textbf{\authorThreeName}}\\
		\footnotesize
		\authorThreeAffil
	\end{textblock}
	\begin{textblock}{\authorCellWidth}[0,0](\colDoubleLeftX,\rowtwoY)
		\centering
		\small{\textbf{\authorFourName}}\\
		\footnotesize
		\authorFourAffil
	\end{textblock}	
	\begin{textblock}{\authorCellWidth}[0,0](\colDoubleRightX,\rowtwoY)
		\centering
		\small{\textbf{\authorFiveName}}\\
		\footnotesize
		\authorFiveAffil
	\end{textblock}	
}
\defcase{6}{
	\begin{textblock}{\authorCellWidth}[0,0](\coloneX,\rowoneY)
		\centering
		\small{\textbf{\authorOneName}}\\
		\footnotesize
		\authorOneAffil
	\end{textblock}	
	\begin{textblock}{\authorCellWidth}[0,0](\coltwoX,\rowoneY)
		\centering
		\small{\textbf{\authorTwoName}}\\
		\footnotesize
		\authorTwoAffil
	\end{textblock}	
	\begin{textblock}{\authorCellWidth}[0,0](\colthreeX,\rowoneY)
		\centering
		\small{\textbf{\authorThreeName}}\\
		\footnotesize
		\authorThreeAffil
	\end{textblock}
	\begin{textblock}{\authorCellWidth}[0,0](\coloneX,\rowtwoY)
		\centering
		\small{\textbf{\authorFourName}}\\
		\footnotesize
		\authorFourAffil
	\end{textblock}	
	\begin{textblock}{\authorCellWidth}[0,0](\coltwoX,\rowtwoY)
		\centering
		\small{\textbf{\authorFiveName}}\\
		\footnotesize
		\authorFiveAffil
	\end{textblock}	
	\begin{textblock}{\authorCellWidth}[0,0](\colthreeX,\rowtwoY)
		\centering
		\small{\textbf{\authorSixName}}\\
		\footnotesize
		\authorSixAffil
	\end{textblock}
}
\defcase{7}{
	\begin{textblock}{\authorCellWidth}[0,0](\coloneX,\rowoneY)
		\centering
		\small{\textbf{\authorOneName}}\\
		\footnotesize
		\authorOneAffil
	\end{textblock}	
	\begin{textblock}{\authorCellWidth}[0,0](\coltwoX,\rowoneY)
		\centering
		\small{\textbf{\authorTwoName}}\\
		\footnotesize
		\authorTwoAffil
	\end{textblock}	
	\begin{textblock}{\authorCellWidth}[0,0](\colthreeX,\rowoneY)
		\centering
		\small{\textbf{\authorThreeName}}\\
		\footnotesize
		\authorThreeAffil
	\end{textblock}
	\begin{textblock}{\authorCellWidth}[0,0](\coloneX,\rowtwoY)
		\centering
		\small{\textbf{\authorFourName}}\\
		\footnotesize
		\authorFourAffil
	\end{textblock}	
	\begin{textblock}{\authorCellWidth}[0,0](\coltwoX,\rowtwoY)
		\centering
		\small{\textbf{\authorFiveName}}\\
		\footnotesize
		\authorFiveAffil
	\end{textblock}	
	\begin{textblock}{\authorCellWidth}[0,0](\colthreeX,\rowtwoY)
		\centering
		\small{\textbf{\authorSixName}}\\
		\footnotesize
		\authorSixAffil
	\end{textblock}
	\begin{textblock}{8}[0,0](\colsinglecentralX,\rowthreeY)
		\centering
		\small{\textbf{\authorSevenName}}\\
		\footnotesize
		\authorSevenAffil
	\end{textblock}
}
\defcase{8}{
	\begin{textblock}{\authorCellWidth}[0,0](\coloneX,\rowoneY)
		\centering
		\small{\textbf{\authorOneName}}\\
		\footnotesize
		\authorOneAffil
	\end{textblock}	
	\begin{textblock}{\authorCellWidth}[0,0](\coltwoX,\rowoneY)
		\centering
		\small{\textbf{\authorTwoName}}\\
		\footnotesize
		\authorTwoAffil
	\end{textblock}	
	\begin{textblock}{\authorCellWidth}[0,0](\colthreeX,\rowoneY)
		\centering
		\small{\textbf{\authorThreeName}}\\
		\footnotesize
		\authorThreeAffil
	\end{textblock}
	\begin{textblock}{\authorCellWidth}[0,0](\coloneX,\rowtwoY)
		\centering
		\small{\textbf{\authorFourName}}\\
		\footnotesize
		\authorFourAffil
	\end{textblock}	
	\begin{textblock}{\authorCellWidth}[0,0](\coltwoX,\rowtwoY)
		\centering
		\small{\textbf{\authorFiveName}}\\
		\footnotesize
		\authorFiveAffil
	\end{textblock}	
	\begin{textblock}{\authorCellWidth}[0,0](\colthreeX,\rowtwoY)
		\centering
		\small{\textbf{\authorSixName}}\\
		\footnotesize
		\authorSixAffil
	\end{textblock}
	\begin{textblock}{\authorCellWidth}[0,0](\colDoubleLeftX,\rowthreeY)
		\centering
		\small{\textbf{\authorSevenName}}\\
		\footnotesize
		\authorSevenAffil
	\end{textblock}
	\begin{textblock}{\authorCellWidth}[0,0](\colDoubleRightX,\rowthreeY)
		\centering
		\small{\textbf{\authorEightName}}\\
		\footnotesize
		\authorEightAffil
	\end{textblock}
}
\defcase{9}{
	\begin{textblock}{\authorCellWidth}[0,0](\coloneX,\rowoneY)
		\centering
		\small{\textbf{\authorOneName}}\\
		\footnotesize
		\authorOneAffil
	\end{textblock}	
	\begin{textblock}{\authorCellWidth}[0,0](\coltwoX,\rowoneY)
		\centering
		\small{\textbf{\authorTwoName}}\\
		\footnotesize
		\authorTwoAffil
	\end{textblock}	
	\begin{textblock}{\authorCellWidth}[0,0](\colthreeX,\rowoneY)
		\centering
		\small{\textbf{\authorThreeName}}\\
		\footnotesize
		\authorThreeAffil
	\end{textblock}
	\begin{textblock}{\authorCellWidth}[0,0](\coloneX,\rowtwoY)
		\centering
		\small{\textbf{\authorFourName}}\\
		\footnotesize
		\authorFourAffil
	\end{textblock}	
	\begin{textblock}{\authorCellWidth}[0,0](\coltwoX,\rowtwoY)
		\centering
		\small{\textbf{\authorFiveName}}\\
		\footnotesize
		\authorFiveAffil
	\end{textblock}	
	\begin{textblock}{\authorCellWidth}[0,0](\colthreeX,\rowtwoY)
		\centering
		\small{\textbf{\authorSixName}}\\
		\footnotesize
		\authorSixAffil
	\end{textblock}
	\begin{textblock}{\authorCellWidth}[0,0](\coloneX,\rowthreeY)
		\centering
		\small{\textbf{\authorSevenName}}\\
		\footnotesize
		\authorSevenAffil
	\end{textblock}	
	\begin{textblock}{\authorCellWidth}[0,0](\coltwoX,\rowthreeY)
		\centering
		\small{\textbf{\authorEightName}}\\
		\footnotesize
		\authorEightAffil
	\end{textblock}	
	\begin{textblock}{\authorCellWidth}[0,0](\colthreeX,\rowthreeY)
		\centering
		\small{\textbf{\authorNineName}}\\
		\footnotesize
		\authorNineAffil
	\end{textblock}
}
\defcase{10}{
	\begin{textblock}{\authorCellWidth}[0,0](\coloneX,\rowoneY)
		\centering
		\small{\textbf{\authorOneName}}\\
		\footnotesize
		\authorOneAffil
	\end{textblock}	
	\begin{textblock}{\authorCellWidth}[0,0](\coltwoX,\rowoneY)
		\centering
		\small{\textbf{\authorTwoName}}\\
		\footnotesize
		\authorTwoAffil
	\end{textblock}	
	\begin{textblock}{\authorCellWidth}[0,0](\colthreeX,\rowoneY)
		\centering
		\small{\textbf{\authorThreeName}}\\
		\footnotesize
		\authorThreeAffil
	\end{textblock}
	\begin{textblock}{\authorCellWidth}[0,0](\coloneX,\rowtwoY)
		\centering
		\small{\textbf{\authorFourName}}\\
		\footnotesize
		\authorFourAffil
	\end{textblock}	
	\begin{textblock}{\authorCellWidth}[0,0](\coltwoX,\rowtwoY)
		\centering
		\small{\textbf{\authorFiveName}}\\
		\footnotesize
		\authorFiveAffil
	\end{textblock}	
	\begin{textblock}{\authorCellWidth}[0,0](\colthreeX,\rowtwoY)
		\centering
		\small{\textbf{\authorSixName}}\\
		\footnotesize
		\authorSixAffil
	\end{textblock}
	\begin{textblock}{\authorCellWidth}[0,0](\coloneX,\rowthreeY)
		\centering
		\small{\textbf{\authorSevenName}}\\
		\footnotesize
		\authorSevenAffil
	\end{textblock}	
	\begin{textblock}{\authorCellWidth}[0,0](\coltwoX,\rowthreeY)
		\centering
		\small{\textbf{\authorEightName}}\\
		\footnotesize
		\authorEightAffil
	\end{textblock}	
	\begin{textblock}{\authorCellWidth}[0,0](\colthreeX,\rowthreeY)
		\centering
		\small{\textbf{\authorNineName}}\\
		\footnotesize
		\authorNineAffil
	\end{textblock}
	\begin{textblock}{8}[0,0](\colsinglecentralX,\rowfourY)
		\centering
		\small{\textbf{\authorTenName}}\\
		\footnotesize
		\authorTenAffil
	\end{textblock}
}

\newcommand\authorOneName{Aleksandar Milenkoski}
\newcommand\authorOneAffil{
	Institute for Program Structures and Data Organization\\
    Karlsruhe Institute of Technology\\
    Karlsruhe, Germany\\
    \emph{milenkoski@kit.edu}\\
}

\newcommand\authorTwoName{Marco Vieira}
\newcommand\authorTwoAffil{
	CISUC, Department of Informatics Engineering\\
	University of Coimbra\\
	Coimbra, Portugal\\
	\emph{mvieira@dei.uc.pt}
}

\newcommand\authorThreeName{Bryan D. Payne}
\newcommand\authorThreeAffil{
	Director, Security Research\\
	Nebula Inc.\\
	Mountain View, CA, USA\\
	\emph{bdpayne@acm.org}\\
}

\newcommand\authorFourName{Nuno Antunes}
\newcommand\authorFourAffil{
	CISUC, Department of Informatics Engineering\\
	University of Coimbra\\
	Coimbra, Portugal\\
	\emph{nmsa@dei.uc.pt}
}

\newcommand\authorFiveName{Samuel Kounev}
\newcommand\authorFiveAffil{
	Chair of Software Engineering\\
    University of W{\"u}rzburg\\
     W{\"u}rzburg, Germany\\
    \emph{samuel.kounev@uni-wuerzburg.de}\\
}

\newcommand\authorSixName{FirstName6 LastName6}
\newcommand\authorSixAffil{
	Department of Cloud Computing3,\\
	State University of SomeCity3,\\
	SomeCity'sLongerName, LongLongLongCountryName4,\\
	e-mail2@e-mail.com
}
\newcommand\authorSevenName{John Doe}
\newcommand\authorSevenAffil{
	Institute of Informatics,\\
	University of Polar Cirlce,\\
	Acity, Acoutry,\\
	john@upc.edu
}
\newcommand\authorEightName{FirstName8 LastName8}
\newcommand\authorEightAffil{
	Department of Cloud Computing3,\\
	State University of SomeCity3,\\
	SomeCity'sLongerName, LongLongLongCountryName4,\\
	e-mail2@e-mail.com
}
\newcommand\authorNineName{FirstName9 LastName9}
\newcommand\authorNineAffil{
	Department of Cloud Computing3,\\
	State University of SomeCity3,\\
	SomeCity'sLongerName, LongLongLongCountryName4,\\
	e-mail2@e-mail.com
}
\newcommand\authorTenName{FirstName10 LastName10}
\newcommand\authorTenAffil{
	Department of Cloud Computing3,\\
	State University of SomeCity3,\\
	SomeCity'sLongerName, LongLongLongCountryName4,\\
	e-mail2@e-mail.com
}

\begin{document}\sloppy
 
\selectlanguage{english} 
\frontmatter

\thispagestyle{empty}
\newcommand{\changefont}[3]{\fontfamily{#1} \fontseries{#2} \fontshape{#3} \selectfont}
\newcommand{\diameter}{20}
\newcommand{\xone}{-25}
\newcommand{\xtwo}{165}
\newcommand{\yone}{20}
\newcommand{\ytwo}{-253}

\newcommand{\rowoneY}{5.5}		
\newcommand{\rowtwoY}{7.0}
\newcommand{\rowthreeY}{8.5}
\newcommand{\rowfourY}{10.1}

\newcommand{\coloneX}{2.5}
\newcommand{\coltwoX}{7.45}
\newcommand{\colthreeX}{12.4}

\newcommand{\colDoubleLeftX}{5}
\newcommand{\colDoubleRightX}{10}

\newcommand{\colsinglecentralX}{5.9}

\newcommand{\authorCellWidth}{4.9}

\begin{titlepage}
\begin{tikzpicture}[overlay]
\draw[color=gray]  
 (\xone mm, \yone mm) -- (\xtwo mm, \yone mm) arc (90:0:\diameter pt) 
  -- (\xtwo mm + \diameter pt , \ytwo mm) -- (\xone mm + \diameter pt , \ytwo mm) 
 arc (270:180:\diameter pt) -- (\xone mm, \yone mm);
\end{tikzpicture}

\changefont{phv}{m}{n}	
\begin{textblock}{14}[0,0](3,2.3)
	\centering
	\large{\TRnumber}\\
	\vspace*{1cm}
	\huge{\mytitle}\\
	\vspace*{0.5cm}
	\Large{\WGname}
\end{textblock}
\begin{textblock}{15.5}[0,0](2,5.2)
	\begin{tikzpicture}
		\fill[red!80!brown] (0,0cm) rectangle (19.5cm,0.1cm);
	\end{tikzpicture}
\end{textblock}

\begin{center}
	\putAuthors{\numAuthors}
\end{center}

\begin{textblock}{14}[0,0](3,13)
	\hfill
	\includegraphics[width=3cm]{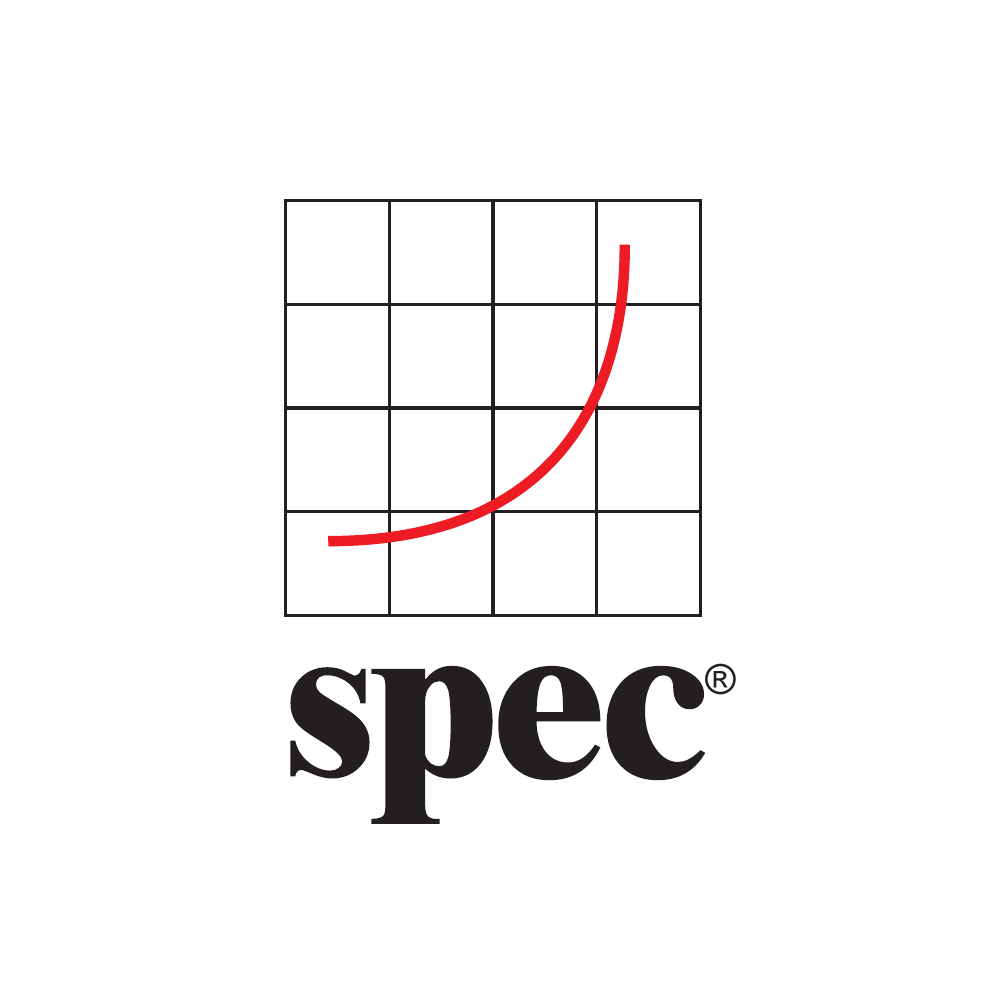} \hfill
	\includegraphics[width=1.9cm]{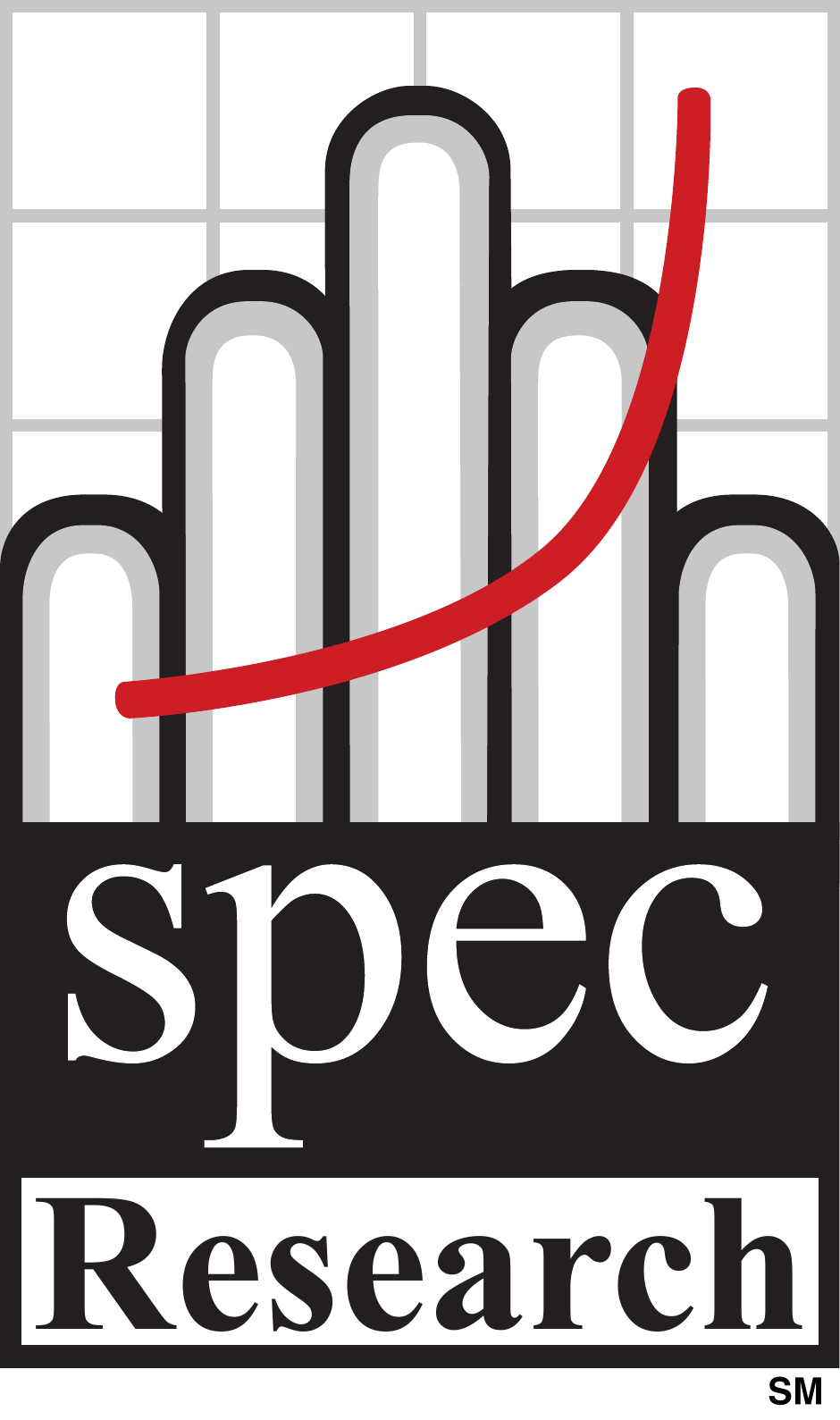} \hspace{1.1cm}\hfill
	\hfill
\end{textblock}

\begin{textblock}{14}[0,0](3,16.75)
	\centering
	\large{\textbf{\TRdate}}
	\hfill
	\large{\textbf{\TRcentralURL}}
	\hfill
	\large{\textbf{\TRrightURL}}
\end{textblock}

\end{titlepage}

\newpage 
\thispagestyle{empty}
\mbox{}

\newpage
\pagenumbering{roman}
\setcounter{tocdepth}{4}
\begin{spacing}{1.5}
 \tableofcontents
 \end{spacing}

\newpage
\thispagestyle{plain}
\section*{Executive Summary}

Modern virtualized service infrastructures expose attack vectors that enable attacks of high severity, such as attacks targeting hypervisors. A malicious user of a guest VM (virtual machine) may execute an attack against the underlying hypervisor via hypercalls, which are software traps from a kernel of a fully or partially paravirtualized guest VM to the hypervisor. The exploitation of a vulnerability of a hypercall handler may have severe consequences such as altering hypervisor's memory, which may result in the execution of malicious code with hypervisor privilege. Despite the importance of vulnerabilities of hypercall handlers, there is not much publicly available information on them. This significantly hinders advances towards securing hypercall interfaces. In this work, we provide in-depth technical information on publicly disclosed vulnerabilities of hypercall handlers. Our vulnerability analysis is based on reverse engineering the released patches fixing the considered vulnerabilities. For each analyzed vulnerability, we provide background information essential for understanding the vulnerability, and information on the vulnerable hypercall handler and the error causing the vulnerability. We also show how the vulnerability can be triggered and discuss the state of the targeted hypervisor after the vulnerability has been triggered.

\vspace{0.5cm}

\textbf{Keywords:\footnote{The keywords used here are defined as part of The 2012 ACM Computing Classification System \cite{acm:classification}.}} \\
Security and Privacy - Systems security  -  Operating systems security - Virtualization and security\\

\mainmatter
\pagenumbering{arabic}
\include{intro}

\section{Introduction}
\label{sec:introduction}

Virtualized environments are becoming increasingly ubiquitous with the growing proliferation of virtualized data centers and cloud environments. However, security concerns are still one of the greatest showstoppers for the wide adoption of cloud computing~\cite{gens:cloud}. Attackers are actively exploring virtualization-specific attack surfaces, such as hypervisors. Attacks targeting hypervisors are of high severity since they may result in crashing the hypervisors including all guest VMs (virtual machines) running on top of them or in altering hypervisors' memory.

A malicious guest VM user may execute an attack against the underlying hypervisor via hypercalls, which are software traps from a kernel of a fully or partially paravirtualized guest VM to the hypervisor. Hypercalls enable intrusion into vulnerable hypervisors initiated from a malicious guest VM kernel. As Rutkowska et al.~\cite{rutkowska:xen} demonstrate, the exploitation of a vulnerability of a hypercall handler (i.e., a hypercall vulnerability) may lead to altering the memory of the targeted hypervisor, which enables, for example, the execution of malicious code with hypervisor privilege. 

Given the severity of attacks triggering hypercall vulnerabilities, the characterization of the hypercall attack surface is a priority since it is crucial for better understanding the security threats posed by hypercall interfaces. The lack of such an understanding significantly hinders advances towards monitoring and securing these interfaces. In-depth technical information on hypercall vulnerabilities is a requirement for characterizing the hypercall attack surface. However, such information is currently very limited. Publicly disclosed vulnerability reports describing hypercall vulnerabilities (e.g., CVE-2013-4494, CVE-2013-3898) are typically the sole source of information and provide only high-level descriptions. There is also no publicly available information on attacks triggering hypercall vulnerabilities performed in practice.

The goal of this work is to provide technical information on hypercall vulnerabilities needed for the improvement of the security of hypercall interfaces (e.g., information on the errors that caused the vulnerabilities and how the vulnerabilities can be triggered). To this end, we analyzed all publicly disclosed hypercall vulnerabilities that we found by searching major CVE (Common Vulnerability and Exposures) report databases (e.g., cvedetails~\cite{cvedet}) based on relevant keywords, such as names of operations of hypercalls. In this work, we focus on the vulnerabilities described in the vulnerability reports  CVE-2012-3494, CVE-2012-3495, CVE-2012-3496, CVE-2012-4539, CVE-2012-5510, CVE-2012-5513, CVE-2012-5525, and CVE-2013-1964. These vulnerabilities are representative of the vulnerabilities that we analyzed in terms of the errors causing them and the ways in which they can be triggered. 
The vulnerabilities considered in this work are from the Xen hypervisor~\cite{xen}, which has the most extensive hypercall interface as opposed to other hypervisors, such as KVM~\cite{kvm}. The considered vulnerabilities are in the handlers of the hypercalls \emph{memory\_op}, \emph{gnttab\_op}, \emph{set\_debugreg}, \emph{physdev\_op}, and \emph{mmuext\_op}.

Our approach for analyzing a hypercall vulnerability consisted of the following steps: \emph{(i)} analysis of the CVE report describing the vulnerability and other relevant information sources, for example, security advisories; \emph{(ii)} reverse engineering of the released patch fixing the vulnerability, and \emph{(iii)} developing proof-of-concept code for triggering the vulnerability. For each considered vulnerability, we provide background information essential for understanding the vulnerability, and information on the vulnerable hypercall handler (i.e., information about the workflow, and input and output  data of the handler) and the error causing the vulnerability. We also show how the vulnerability can be triggered and discuss the state of the targeted hypervisor after the vulnerability has been triggered. 

We stress that we provide information on a vulnerable hypercall handler to the extent that is relevant for understanding a given vulnerability, for example, we discuss only some input parameters of the handler. We also stress that we do not provide proof-of-concept code for triggering the considered vulnerabilities ready for use. We present only the hypercalls executed as part of an attack triggering a given hypercall vulnerability, and the values of relevant hypercall parameters (i.e., parameters identifying the executed hypercalls and, where applicable, parameters with values specifically crafted for triggering the vulnerability). Finally, we stress that we do not demonstrate vulnerability exploitation where it is possible (e.g., malicious code execution). We focus instead on the errors causing the considered vulnerabilities, the activities for triggering them, and the effects of triggering the vulnerabilities on the state of the vulnerable hypervisors. We argue that the information that we provide is relevant for better understanding the security threats that hypercall interfaces pose, which will help to focus approaches
for improving the security of hypervisors.

\section{Information on hypercall vulnerabilities}
\label{sec:information}

\subsection{Hypercall memory\_op}

The \emph{memory\_op} hypercall is used for managing the memory of a guest VM, for example, altering the layout of a given memory region.\footnote{We refer the reader to~\cite{xenmanual} for more information on the functionalities of the \emph{memory\_op} hypercall.}  In the handler of \emph{memory\_op}, the different types of memory addresses that the Xen hypervisor supports for abstracting physical memory available to guest VMs are used for accessing locations in memory:

\begin{quote}
\begin{compactitem}[$\circ$]
\item \emph{virtual address} - an address of a location in the virtual memory of a guest VM; 
\item \emph{GPFN (Guest Pseudo-Physical Frame Number)} - an address of a page frame that is a physical memory address from the perspective of a guest VM; 
\item \emph{GMFN (Guest Machine Frame Number)} - an address of a page frame that is a machine address from the perspective of a guest VM;
\item \emph{MFN (Machine Frame Number)} - an address of a page frame that is a real machine address. 
\end{compactitem}
\end{quote}

For accessing contiguous memory blocks, the different types of addresses mentioned above are used for accessing \emph{extents} of a given \emph{order} such that an extent consists of $2^{order}$ memory pages. 

Mappings between the different types of memory addresses are stored in tables for that purpose. Mappings between virtual addresses and GPFNs are stored in a page table, between GPFNs and GMFNs in a physical-to-machine table, and between GMFNs and GPFNs in a machine-to-physical table. 

We refer to reader to~\cite{xenwiki} and~\cite{chisnall:thedefinitive} for further information on how the Xen hypervisor manages memory.  

\paragraph{Vulnerability CVE-2012-3496} 

\begin{quote}
``XENMEM\_populate\_physmap in Xen 4.0, 4.1, and 4.2, and Citrix XenServer 6.0.2 and earlier, when translating paging mode is not used, allows local PV OS guest kernels to cause a denial of service (BUG triggered and host crash) via invalid flags such as MEMF\_populate\_on\_demand."~\cite{cve20123496}
\end{quote}

\emph{XENMEM\_populate\_physmap} is an operation of the \emph{memory\_op} hypercall, which is used for requesting extents from the hypervisor. \emph{XENMEM\_populate\_physmap} is also used for marking extents as ``populate-on-demand''. Extents marked as ``populate-on-demand'' can be assigned to the physical memory of a given guest VM, or removed from it, on demand at run time. 

\smallskip
\textbf{Input:}\footnote{As in Xen of version 4.1.0.\label{fnt:2012346}} \emph{XENMEM\_populate\_physmap} takes as input a structure of type \emph{xen\_memory\_reservation}. which is defined as: 

\begin{lstlisting}[breaklines=true, backgroundcolor=\color{light-gray},escapeinside=`', basicstyle=\small, frame=tb, columns=fullflexible, ]
struct xen_memory_reservation {
  GUEST_HANDLE(xen_pfn_t)	extent_start;
  unsigned int		extent_order;
  unsigned int		address_bits;
  . . .
}
\end{lstlisting}

\emph{extent\_start} stores the virtual address of the head of an array that contains memory addresses (GPFNs) at which the extents obtained from the hypervisor are to be mapped, or addresses (GPFNs) of the beginnings of the extents that are to be marked as ``populate-on-demand''; \emph{extent\_order} stores the order of a single extent; \emph{address\_bits} stores the flags of the \emph{XENMEM\_populate\_physmap} hypercall operation, one of which is \emph{MEMF\_populate\_on\_demand}. \emph{MEMF\_populate\_on\_demand} is enabled when \emph{XENMEM\_populate\_physmap} is used for marking extents as ``populate-on-demand''.

\smallskip
\textbf{Output:}\textsuperscript{\ref{fnt:2012346}} On success, \emph{XENMEM\_populate\_physmap} returns the number of the obtained extents or of the extents marked as ``populate-on-demand''. In case \emph{XENMEM\_populate\_physmap} has been used for obtaining extents, the array that starts at the virtual address stored in \emph{extent\_start} is populated with the memory addresses (MFNs) of the beginnings of the obtained extents. On failure, \emph{XENMEM\_populate\_physmap} returns an error code (typically a negative integer value). 
 
 \smallskip
\textbf{Workflow of the vulnerable hypercall handler:}\textsuperscript{\ref{fnt:2012346}} 
\begin{lstlisting}[breaklines=true, backgroundcolor=\color{light-gray},escapeinside=`', basicstyle=\small, frame=tb, columns=fullflexible, ]
do_memory_op (XENMEM_populate_physmap, (struct xen_memory_reservation) res)
  . . .
  `\textbf{call}' populate_physmap(...) 
    . . .
      `\textbf{for each}' GPFN in res.extent_start
        `\textbf{if}' MEMF_populate_on_demand
          `\textbf{call}' guest_physmap_mark_populate_on_demand(...)
            `\textit{\textbf{call BUG\_ON(...)}}'  
            . . .
          `\textbf{return}' 
          . . .
        `\textbf{else}':
          . . .
    . . . 
  `\textbf{return}' 
`\textbf{return}' 
\end{lstlisting}

\smallskip
\textbf{Description of the vulnerability:} In \emph{guest\_physmap\_mark\_populate\_on\_demand}, a function invoked in the handler of \emph{XENMEM\_populate\_physmap}, the BUG\_ON macro is used for checking whether the guest VM from where the \emph{memory\_op} hypercall has been invoked has the ``translated paging'' mode disabled. BUG\_ON is a macro that crashes the system where it is executed if the condition that it evaluates is true. If \emph{guest\_physmap\_mark\_populate\_on\_demand} is invoked from a paravirtualized guest VM (note that paravirtualized guest VMs have the ``translated paging'' mode disabled by default), the condition that the BUG\_ON macro evaluates is true and the hypervisor crashes. Thus, CVE-2012-3496 can be triggered by invoking the \emph{XENMEM\_populate\_physmap} hypercall operation, with the \emph{MEMF\_populate\_on\_demand} flag enabled, from a paravirtualized guest VM.

\smallskip
\textbf{Vulnerability fix:} A patch fixing the vulnerability CVE-2012-3496 was released on 5 September 2012 and is available at~\cite{patch20123496}. The patch replaces the \emph{BUG\_ON} macro with an \emph{if} clause. 

\smallskip
\textbf{Triggering CVE-2012-3496:} We triggered CVE-2012-3496 in the following environment:
\begin{quote}
\begin{compactitem}[$\circ$]
\item guest VM: OS - Debian Squeeze (64 bit), kernel - 2.6.32-5-amd64; 
\item host VM: OS - Debian Squeeze (64 bit), kernel - 2.6.32-5-amd64;
\item hypervisor: Xen 4.1.0.
\end{compactitem}
\end{quote}

\begin{figure}[ht]
\centering
\includegraphics[scale=1.0]{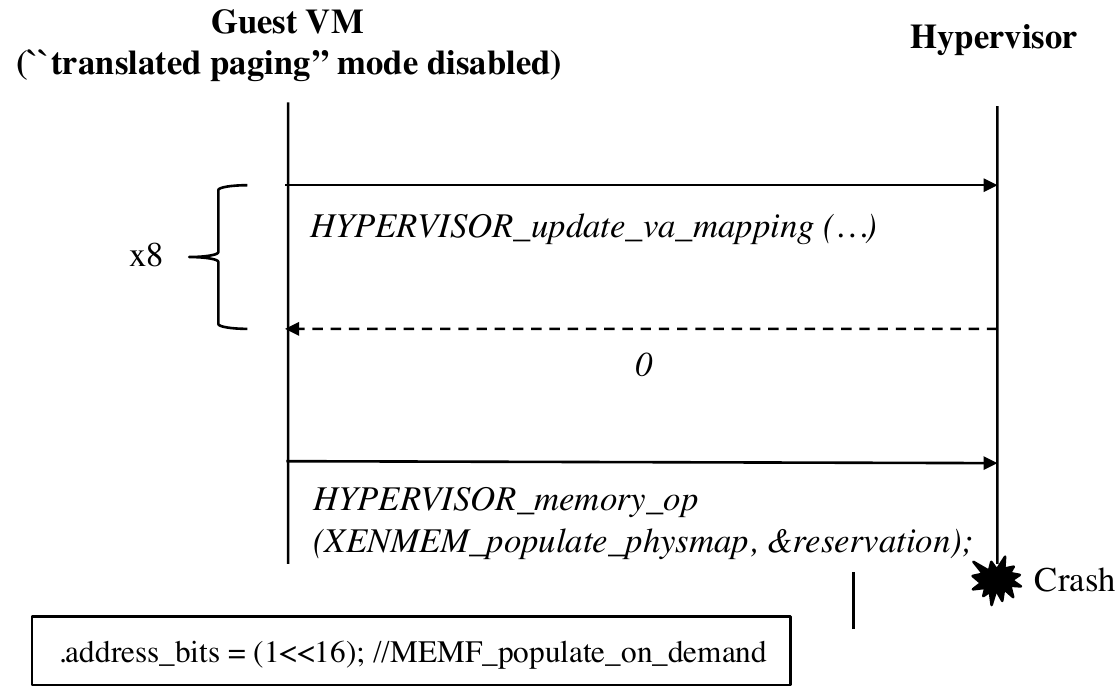}
\caption{An attack triggering CVE-2012-3496}
\label{fig:cve20123496}
\end{figure}

The attack that we executed is depicted in Figure~\ref{fig:cve20123496}.

\smallskip
\textbf{Post-attack state of the hypervisor:} The hypervisor crashes when the BUG\_ON macro is executed.

\paragraph{Vulnerability CVE-2012-5513} 

\begin{quote}
``The XENMEM exchange handler in Xen 4.2 and earlier does not properly check the memory address, which allows local PV guest OS administrators to cause a denial of service (crash) or possibly gain privileges via unspecified vectors that overwrite memory in the hypervisor reserved range.''~\cite{cve20125513}
\end{quote}

\emph{XENMEM\_exchange} is an operation of the \emph{memory\_op} hypercall, which is used for modifying the layout of a memory region of a guest VM by ``exchanging'' extents between the guest VM and the hypervisor. The latter is performed by remapping a set of memory addresses (GPFNs) of beginnings of extents of the guest VM to memory addresses (GMFNs) of beginnings of extents, requested by the guest VM and allocated by the hypervisor for the ``exchange'' operation. For instance, \emph{XENMEM\_exchange} can be used for defragmenting memory such that, for example, 2 extents consisting of 2 pages are exchanged for a single extent consisting of 4 pages.

\smallskip
\textbf{Input:}\footnote{As in Xen of version 4.1.0.\label{fnt:20125513}} \emph{XENMEM\_exchange} takes as input a structure of type \emph{xen\_memory\_exchange} defined as:

\begin{lstlisting}[breaklines=true, backgroundcolor=\color{light-gray},escapeinside=`', basicstyle=\small, frame=tb, columns=fullflexible, ]
struct xen_memory_exchange {
  struct xen_memory_reservation in;
  struct xen_memory_reservation out;
  xen_ulong_t nr_exchanged;
}
\end{lstlisting}

, where \emph{xen\_memory\_reservation} is defined as:

\begin{lstlisting}[breaklines=true, backgroundcolor=\color{light-gray},escapeinside=`', basicstyle=\small, frame=tb, columns=fullflexible, ]
struct xen_memory_reservation {
  GUEST_HANDLE(xen_pfn_t) extent_start;
  unsigned int extent_order;
  xen_ulong_t nr_extents;
  . . .
}
\end{lstlisting}

The fields of the \emph{(struct xen\_memory\_exchange) in} structure store information about the extents that are to be ``exchanged''. \emph{in.nr\_extents} stores the number of extents to be ``exchanged''; \emph{in.extent\_start} stores the virtual address of the head of an array that contains the memory addresses (GMFNs) of the beginnings of the extents to be ``exchanged''; \emph{in.extent\_order} stores the order of a single extent. 

The fields of the \emph{(struct xen\_memory\_exchange) out} structure store information about the extents requested from the hypervisor. \emph{out.nr\_extents} stores the number of requested extents; \emph{out.extent\_order} stores the order of a single requested extent; \emph{out.extent\_start} stores the virtual address of the head of an array that consists of GPFNs at which the requested extents are to be mapped in guest VM's memory.

\smallskip
\textbf{Output:}\textsuperscript{\ref{fnt:20125513}} On success, \emph{XENMEM\_exchange} returns 0. The array that starts at the address stored in \emph{(struct xen\_memory\_exchange) out.extent\_start} is populated with the memory addresses (GMFNs) of the beginnings of the extents allocated by the hypervisor for the ``exchange'' operation. On failure, \emph{XENMEM\_exchange}  returns an error code (typically a negative integer value). 

\smallskip
\textbf{Workflow of the vulnerable hypercall handler:}\textsuperscript{\ref{fnt:20125513}}
\begin{lstlisting}[breaklines=true, backgroundcolor=\color{light-gray},escapeinside=`', basicstyle=\small, frame=tb,columns=fullflexible]
do_memory_op (XENMEM_exchange, (struct xen_memory_exchange) exch)
  `\textbf{call}' memory_exchange (XENMEM_exchange, (struct xen_memory_exchange) exch)
    . . .
    allocate extent(s) of `2$^{exch.in.order}$' pages
    store the addresses (GMFNs) of the beginnings of the allocated extents in array `\emph{mfn}'    
    . . .    
    `\textit{\textbf{call  \_\_copy\_to\_guest\_offset(...)}}'
         populate memory beginning at `\emph{exch.out.extent\_start}' with the GMFNs in `\emph{mfn}'
    `\textbf{return}' 
    . . .
  `\textbf{return}' 
`\textbf{return}' 
\end{lstlisting}

\smallskip
\textbf{Description of the vulnerability:} The function \emph{\_\_copy\_to\_guest\_offset(to, offset, from, size)}, which is invoked in the handler of the \emph{XENMEM\_exchange} hypercall operation, copies data from a virtual address in hypervisor context  (\emph{from}) to a virtual address in guest VM context (\emph{to}). For the sake of performance, \emph{\_\_copy\_to\_guest\_offset(to, offset, from, size)} did not perform value validation of the \emph{from} and \emph{to} parameters. As a result, a malicious VM user can invoke \emph{\_\_copy\_to\_guest\_offset(to, offset, from, size)} such that \emph{to} is an address reserved for use by the hypervisor, which leads to overwriting hypervisor's memory. CVE-2012-5513 can be triggered by invoking the \emph{XENMEM\_exchange} hypercall operation with an address reserved for use by the hypervisor stored in the \emph{(struct xen\_memory\_exchange) out.extent\_start} parameter.

\smallskip
\textbf{Vulnerability fix:} A patch fixing the vulnerability CVE-2012-5513 was released on 3 December 2012 and is available at~\cite{patch20125513}. The patch inserts an invocation of the function \emph{guest\_handle\_okay} in the handler of the \emph{XENMEM\_exchange} hypercall operation, which validates the values of the \emph{from} and \emph{to} parameters of  \emph{\_\_copy\_to\_guest\_offset}. For instance, a valid virtual address is an address that is not reserved for use by the hypervisor. 

\smallskip
\textbf{Triggering CVE-2012-5513:} We triggered CVE-2012-5513 in the following environment:

\begin{quote}
\begin{compactitem}[$\circ$]
\item guest VM - OS: Debian Squeeze (64 bit), kernel 2.6.32-5-amd64; 
\item host VM - OS: Debian Squeeze (64 bit), kernel 2.6.32-5-amd64; 
\item hypervisor - Xen 4.1.0. 
\end{compactitem}
\end{quote}

\begin{figure}[t]
\centering
\includegraphics[scale=1.0]{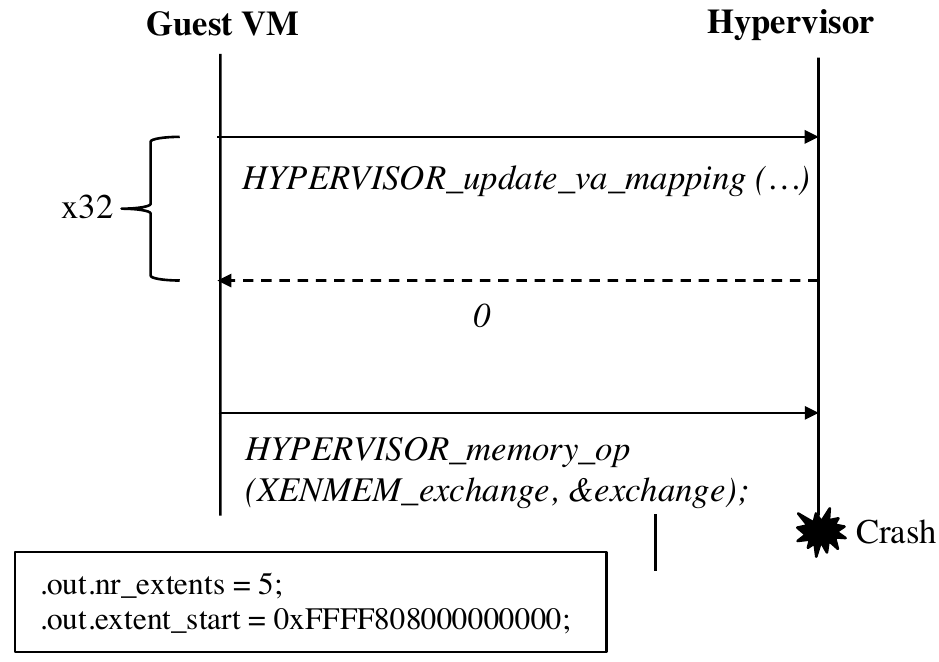}
\caption{An attack triggering CVE-2012-5513}
\label{fig:cve20125513}
\end{figure}

The attack that we executed is depicted in Figure~\ref{fig:cve20125513}.

 \smallskip
\textbf{Post-attack state of the hypervisor:} When CVE-2012-5513 is triggered, the memory region of the hypervisor beginning at the address stored in \emph{(struct xen\_memory\_exchange) out.extent\_start} is overwritten with the memory addresses (GMFNs) of the beginnings of the extents allocated by the hypervisor for the ``exchange'' operation. Thus, an attacker cannot control the values with which the hypervisor's memory is overwritten. The amount of data written to the hypervisor's memory is \emph{(struct xen\_memory\_exchange) out.nr\_extents} bytes. 

Triggering CVE-2012-5513 may result in a crash of the hypervisor or corrupting its state.  Whether the hypervisor crashes depends on which region of the hypervisor's memory is overwritten. An attacker can specify a memory region for overwriting by storing values in the parameters \emph{(struct xen\_memory\_exchange) out.extent\_start} and \emph{(struct xen\_memory\_exchange) out.nr\_extents}. For instance, when we triggered CVE-2012-5513 in our testbed environment, for the values of 0xFFFF808000000000 and 32, and 0xFFFF808000000000 and 16, of \emph{(struct xen\_memory\_exchange) out.extent\_start} and \emph{(struct xen\_memory\_exchange) out.nr\_extents}, respectively, the hypervisor crashed. For the values of 0xFFFF808000000000 and 8 of \emph{(struct xen\_memory\_exchange) out.extent\_start} and \emph{(struct xen\_memory\_exchange) out.nr\_extents}, the hypervisor continued operating with its memory overwritten. 
 
\subsection{Hypercall gnttab\_op}

The \emph{gnttab\_op} hypercall is used for managing grant tables. \emph{Grant tables} provide a mechanism for sharing memory between guest VMs (\emph{domains} in Xen terminology) running on top of a Xen hypervisor; that is, it enables the sharing of page frames by granting page frame access permissions to domains or transferring ownerships of pages between domains. Each domain maintains a grant table, which is shared with the hypervisor. A grant table consists of \emph{grant table entries} (i.e., \emph{grants}) indexed by grant references (i.e., \emph{grefs}). In order to access a page frame for which it needs an access permission, a domain first has to \emph{acquire} the grant that grants the access permission from the domain that has issued the grant. When an acquired grant is not needed anymore, it is \emph{released}.

There are version 1 and version 2 grant tables.  The format of a grant table entry of a version 1 grant table is \emph{[gref][domid][frame][flags]}, where \emph{gref} is a grant reference, \emph{domid} is the identification number of domain to which permissions are granted, \emph{frame} is the MFN of the page frame for which permissions are granted, and \emph{flags} are the permissions granted (e.g., read, write, or read and write permissions), which are also referred to as \emph{status} of a grant table entry. 

Grant tables of version 2, in addition to grants of the format mentioned above, support transitive grants. \emph{Transitive grants} are used for granting transitive permissions such that a domain issues a grant that refers to a grant issued by another domain.

For the sake of performance, the status of grant table entries of a grant table of version 2 are stored in \emph{status frames}, which are separate from the frames where the rest of the grant table entries are stored. 

There are shared and active grants. \emph{Shared grants} are grants issued by a domain. \emph{Active grants} are grants that are in use (i.e., that are acquired) at a given time. A transitive active grant has the fields \emph{trans\_domain} and \emph{trans\_gref}, where \emph{trans\_domain} is the domain that has issued the grant to which the transitive grant refers, and \emph{trans\_gref} is the reference of the grant to which the transitive grant refers.

For in-depth information on the grant table mechanism of the Xen hypervisor we refer the reader to~\cite{xenwiki} and~\cite{chisnall:thedefinitive}.

\paragraph{Vulnerability CVE-2012-4539} 

\begin{quote}
``Xen 4.0 through 4.2, when running 32-bit x86 PV guests on 64-bit hypervisors, allows local guest OS administrators to cause a denial of service (infinite loop and hang or crash) via invalid arguments to GNTTABOP\_get\_status\_frames, aka Grant table hypercall infinite loop DoS vulnerability.''~\cite{cve20124539}
\end{quote}

\emph{GNTTABOP\_get\_status\_frames} is an operation of the \emph{grant\_table\_op} hypercall, which is used for retrieving MFNs of status frames (i.e., status frame MFNs) of a domain. 

\smallskip
\textbf{Input:}\footnote{As in Xen of version 4.1.2.\label{fnt:20124539}}  \emph{GNTTABOP\_get\_status\_frames} takes as input a structure of type \emph{gnttab\_get\_status\_frames} defined as: 

\begin{lstlisting}[breaklines=true, backgroundcolor=\color{light-gray},escapeinside=`', basicstyle=\small, frame=tb, columns=fullflexible, ]
struct gnttab_get_status_frames {
  uint32_t nr_frames;
  domid_t  dom;
  int16_t  status;              
  XEN_GUEST_HANDLE(uint64_t) frame_list;
}
\end{lstlisting}

\emph{nr\_frames} stores the number of requested status frame MFNs; \emph{dom} stores the identification number of the domain whose status frame MFNs are requested; \emph{frame\_list} stores the virtual address of the head of an array where status frame MFNs are to be stored upon successful completion of the \emph{GNTTABOP\_get\_status\_frames} operation.

\smallskip
\textbf{Output:}\textsuperscript{\ref{fnt:20124539}} On success, a return code is stored in \emph{(struct gnttab\_get\_status\_frames) status} and the list starting at the address stored in \emph{struct gnttab\_get\_status\_frames) frame\_list} is populated with status frame MFNs. On failure, \emph{XENMEM\_populate\_physmap} returns an error code (typically a negative integer value). 

\smallskip
\textbf{Workflow of the vulnerable hypercall handler:}\textsuperscript{\ref{fnt:20124539}}
\begin{lstlisting}[breaklines=true, backgroundcolor=\color{light-gray},escapeinside=`', basicstyle=\small, frame=tb,columns=fullflexible]
compat_grant_table_op(GNTTABOP_get_status_frames, (struct gnttab_get_status_frames) gf, int count = 1)
  rc = 0
  i = 0
  `\textbf{for}' `\textit{\textbf{i < count}}' and rc = 0
    . . .
    `\textbf{if}' count = 1
        `\textbf{call}' gnttab_get_status_frames(gf, ...)
           . . .
           `\textbf{if}' gf.nr_frames > the number of status frames of domain gf.dom
              `\textit{\textbf{gf.status = GNTST\_general\_error}}' 
           `\textbf{else}'
              . . .
              gf.status = GNTST_okay
        `\textbf{return}'
        `\textbf{if}' gf.status = GNTST_okay 
           increment i to gf.nr_frames
        . . .
`\textbf{return}'
\end{lstlisting}

\smallskip
\textbf{Description of the vulnerability:} In the hypercall handler \emph{compat\_grant\_table\_op}, a \emph{for} cycle loops until the value of the variable \emph{i}, which is initialized to 0, is smaller than the value of the input parameter \emph{count}, which has to be 1. In \emph{compat\_grant\_table\_op}, the value of the variable \emph{i} is incremented to the value of the input parameter \emph{(struct gnttab\_get\_status\_frames) nr\_frames} only if \emph{(struct gnttab\_get\_status\_frames) status} stores the value of the constant variable \emph{GNTST\_okay}. The value of \emph{(struct gnttab\_get\_status\_frames) status} is set in the function \emph{gnttab\_get\_status\_frames}, which is invoked in \emph{compat\_grant\_table\_op}. \emph{gnttab\_get\_status\_frames} sets the value of \emph{(struct gnttab\_get\_status\_frames) status} to the value of \emph{GNTST\_okay} only if the value of the input parameter \emph{(struct gnttab\_get\_status\_frames) nr\_frames} is smaller than the number of status frames of the domain whose identification number is stored in the parameter \emph{(struct gnttab\_get\_status\_frames) dom}. 

CVE-2012-4539 can be triggered by invoking \emph{GNTTABOP\_get\_status\allowbreak\_frames} such that the value of the input parameter \emph{(struct gnttab\_get\_status\_frames) nr\_frames} is greater than the number of status frames of the domain whose identification number is stored in the parameter \emph{(struct gnttab\_get\_status\_frames) dom}. This results in infinite looping of the \emph{for} cycle in \emph{compat\_grant\_table\_op}. 

In order to trigger CVE-2012-4539, one has to set the value of \emph{(struct gnttab\_get\_status\_frames) nr\_frames} to a value greater than $\lceil \frac{nr\_grants \times  sizeof(uint16\_t) }{PAGE\_SIZE} \rceil $, where \emph{nr\_grants} is the number of grants issued by the domain whose identification number is stored in \emph{(struct gnttab\_get\_status\_frames) dom}, \emph{PAGE\_SIZE} is the size of a single page of the domain, and \emph{uint16\_t} is the size of a variable of type unsigned 16-bit integer. 

Since the erroneous code is in the handler \emph{compat\_grant\_table\_op}, CVE-2012-4539 can be triggered only from a 64-bit guest VM running on top of a 32-bit host VM. 

\smallskip
\textbf{Vulnerability fix:} A patch fixing the vulnerability CVE-2012-4539 was released on 13 November 2012 and is available at~\cite{patch20124539}. The patch modifies \emph{compat\_grant\_table\_op} such that the value of \emph{i} is set to 1, which is equal to the value of \emph{count}, if the value of \emph{(struct gnttab\_get\_status\_frames) status} is not equal to the value of \emph{GNTST\_okay}. This prevents the \emph{for} cycle in \emph{compat\_grant\_table\_op} from looping indefinitely. 

\smallskip
\textbf{Triggering CVE-2012-4539:} We triggered CVE-2012-4539 in the following environment:

\begin{quote}
\begin{compactitem}[$\circ$]
\item guest VM - OS: Ubuntu Precise (32 bit), kernel 3.8.0-29-generic; 
\item host VM - OS: Ubuntu Precise (64 bit), kernel 3.8.0-29-generic; 
\item hypervisor - Xen 4.1.2.
\end{compactitem}
\end{quote}

\begin{figure}[ht]
\centering
\includegraphics[scale=1.0]{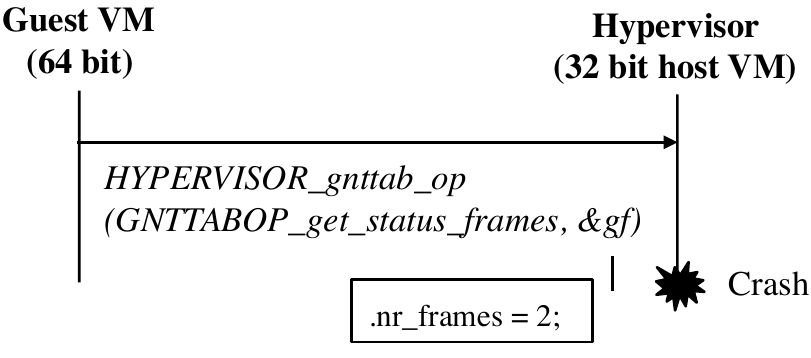}
\caption{An attack triggering CVE-2012-4539}
\label{fig:cve20124539}
\end{figure}

The attack that we executed is depicted in Figure~\ref{fig:cve20124539}.

\smallskip
\textbf{Post-attack state of the hypervisor:} When we triggered CVE-2012-4539 in our testbed environment, the guest VM from where we invoked \emph{GNTTABOP\_get\_status\_frames} hanged. When we issued the \emph{xm/xl destroy} command to shutdown the non-responsive guest VM, the hypervisor crashed. The hypervisor did not crash when we issued the \emph{xm/xl shutdown} command to shutdown, and the \emph{xm/xl reboot} command to reboot, the non-responsive guest VM.

\paragraph{Vulnerability CVE-2012-5510} 

\begin{quote}

``Xen 4.x, when downgrading the grant table version, does not properly remove the status page from the tracking list when freeing the page, which allows local guest OS administrators to cause a denial of service (hypervisor crash) via unspecified vectors.''~\cite{cve20125510}
\end{quote}

The \emph{GNTTABOP\_set\_version} is an operation of the \emph{grant\_table\_op} hypercall, which is used for downgrading (from version 2 to version 1) or upgrading (from version 1 to version 2) grant tables. 

\smallskip
\textbf{Input:}\footnote{As in Xen of version 4.1.2.\label{fnt:20125510}} \emph{GNTTABOP\_set\_version} takes as input a structure of type \emph{gnttab\_set\_version} defined as: 

\begin{lstlisting}[breaklines=true, backgroundcolor=\color{light-gray},escapeinside=`', basicstyle=\small, frame=tb, columns=fullflexible, ]
struct gnttab_set_version {
  uint32_t version;
}
\end{lstlisting}

\emph{version} stores the version to which the grant table of the domain from where \emph{GNTTABOP\_set\_version} is invoked is to be set. 

\smallskip
\textbf{Output:}\textsuperscript{\ref{fnt:20125510}} On success, \emph{GNTTABOP\_set\_version} returns 0 and the version of the grant table from where \emph{GNTTABOP\_set\_version} has been invoked is stored in \emph{(struct gnttab\_set\_version) version}. On failure,  \emph{GNTTABOP\_set\_version} returns an error code (typically a negative integer value). 

\smallskip
\textbf{Workflow of the vulnerable hypercall handler:}\textsuperscript{\ref{fnt:20125510}}

\begin{lstlisting}[breaklines=true, backgroundcolor=\color{light-gray},escapeinside=`', basicstyle=\small, frame=tb,columns=fullflexible]
do_grant_table_op(GNTTABOP_set_version, ...)
  `\textbf{call}' gnttab_set_version(...)
    . . .
      `\textbf{if}' upgrading grant table
        `\textbf{call}' gnttab_populate_status_frames(...)
            allocate status frames 
        `\textbf{return}'
      `\textbf{if}' downgrading grant table 
        `\textit{\textbf{call gnttab\_unpopulate\_status\_frames(...)}}'
            `\textit{\textbf{release status frames}}'
        `\textbf{return}'
    . . .     
  `\textbf{return}'
`\textbf{return}'
\end{lstlisting}

\smallskip
\textbf{Description of the vulnerability:} The function \emph{gnttab\_unpopulate\_status\_frames}, which is invoked in the handler of the \emph{GNTTABOP\_set\_version} hypercall operation, releases allocated status frames when a grant table is downgraded. However, \emph{gnttab\_unpopulate\_status\_frames} does not fully perform the procedure for releasing status frames; that is, it does not remove the nodes that are associated with the status frames being released from the \emph{xenpage\_list} linked list. \emph{xenpage\_list} is a list of nodes that contain information about frames allocated from the hypervisor's heap memory space for the needs of a given guest VM. 

Since \emph{gnttab\_unpopulate\_status\_frames} does not remove from \emph{xenpage\_list} the nodes associated with the status frames, subsequent allocation of the same frames leads to adding nodes to \emph{xenpage\_list} that are duplicates of the nodes that have not been removed by \emph{gnttab\_unpopulate\allowbreak\_status\_frames}. This is effectively a corruption of \emph{xenpage\_list}. The \emph{gnttab\_populate\_status\_frames} function, which is invoked in the handler of \emph{GNTTABOP\_set\_version} when a grant table is upgraded, may be used for allocating the same frames that have been released when a grant table has been downgraded.

CVE-2012-5510 can be triggered by continuously allocating and releasing status frames, which eventually leads to corruption of \emph{xenpage\_list}; that is, CVE-2012-5510 can be triggered by continuously upgrading and downgrading a grant table. When a corruption of \emph{xenpage\_list} occurs depends on the amount of free heap memory of the targeted hypervisor as well as the memory allocating mechanism used. 

\smallskip
\textbf{Vulnerability fix:} A patch fixing the vulnerability CVE-2012-5510 was released on 3 December 2012 and is available at~\cite{patch20125510}. The patch modifies the function \emph{gnttab\_unpopulate\_status\_frames} such that it inserts an invocation of the function \emph{put\_page}. \emph{put\_page} removes from \emph{xenpage\_list} the nodes associated with the status frames being released when a grant table is downgraded.

\smallskip
\textbf{Triggering CVE-2012-5510:} We triggered CVE-2012-5510 in the following environment:

\begin{quote}
\begin{compactitem}[$\circ$]
\item guest VM - OS: Ubuntu Precise (32 bit), kernel 3.8.0-29-generic; 
\item host VM - OS: Ubuntu Precise (32 bit), kernel 3.8.0-29-generic; 
\item hypervisor - Xen 4.1.2.
\end{compactitem}
\end{quote}

\begin{figure}[ht]
\centering
\includegraphics[scale=1.0]{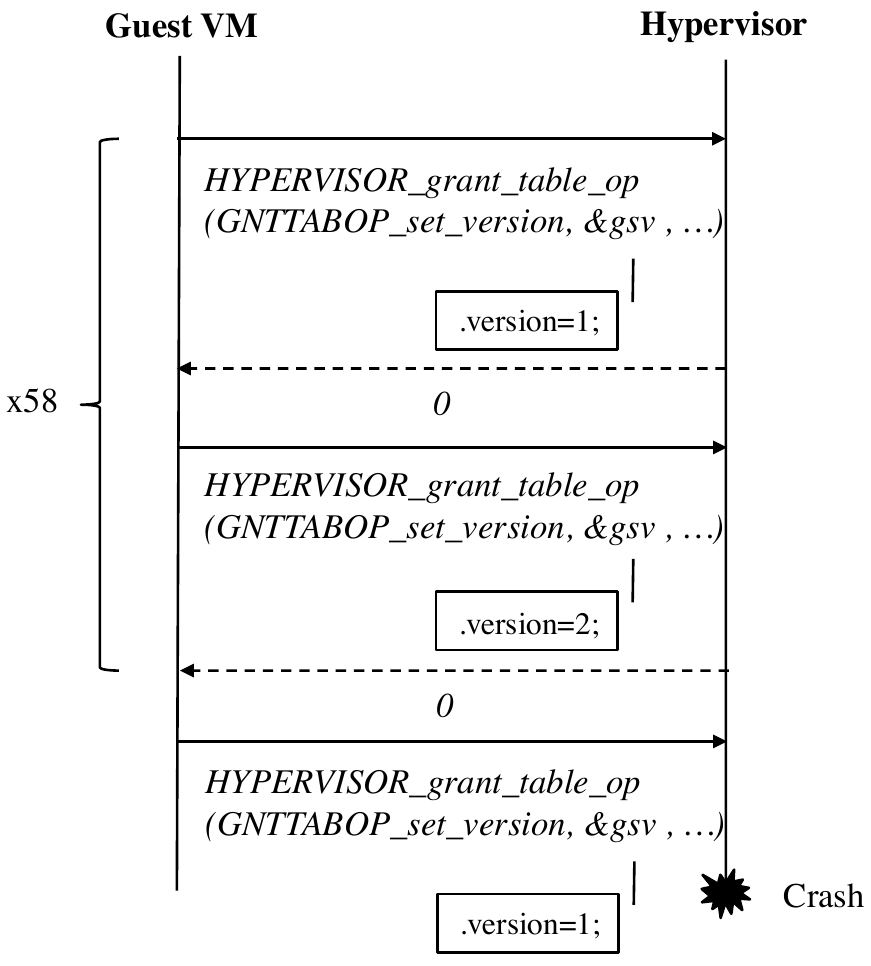}
\caption{An attack triggering CVE-2012-5510}
\label{fig:cve20125510}
\end{figure}

The attack that we executed is depicted in Figure~\ref{fig:cve20125510}.

\smallskip
\textbf{Post-attack state of the hypervisor:} Depending on the use of \emph{xenpage\_list} after it has been corrupted, triggering CVE-2012-5510 may result in crash of the targeted hypervisor or may corrupt its state. The hypervisor crashed when we triggered CVE-2012-5510 in our testbed environment.

\paragraph{Vulnerability CVE-2013-1964} 

\begin{quote}
``Xen 4.0.x and 4.1.x incorrectly releases a grant reference when releasing a non-v1, non-transitive grant, which allows local guest administrators to cause a denial of service (host crash), obtain sensitive information, or possible have other impacts via unspecified vectors.''~\cite{cve20131964}
\end{quote}

\emph{GNTTABOP\_copy} is an operation of the \emph{grant\_table\_op} hypercall, which is used for copying memory pages from a source domain (SD) (i.e., the domain to which the page being copied is allocated) to a destination domain (DD) (i.e., the domain to which the page is copied) with respect to the data read and write permissions set by the SD and/or the DD using grant tables. \emph{GNTTABOP\_copy} can be invoked from the SD, the DD, or a domain that is neither the SD or the DD. The domain from where \emph{GNTTABOP\_copy} is invoked is called the \emph{local} domain, whereas the other domains involved in copying pages are called \emph{remote} domains. 

\smallskip
\textbf{Input:}\footnote{As in Xen of version 4.1.2.\label{fnt:20131964}}  GNTTABOP\_copy takes as input a structure of type \emph{gnttab\_copy} defined as:

\begin{lstlisting}[breaklines=true, backgroundcolor=\color{light-gray},escapeinside=`', basicstyle=\small, frame=tb, columns=fullflexible, ]
struct gnttab_copy {
  struct {
    union {
      grant_ref_t ref;
      xen_pfn_t   gmfn;
    } u;
    domid_t  domid;
    . . .
  } source, dest;
  uint16_t      len;
  uint16_t      flags;          
  int16_t       status;
}
\end{lstlisting}

\emph{source.u.gmfn} stores the GMFN of the page that is to be copied if the SD is a local domain; \emph{dest.u.gmfn} stores the GMFN of the page of the DD to which a page of the SD is to be copied if the DD is a local domain; \emph{source.u.ref} stores the grant reference of the grant that grants access to the page that is to be copied if the SD is a remote domain; \emph{dest.u.ref} stores the grant reference of the grant that grants access to the page of the DD to which a page from the SD is to be copied in case the DD is a remote domain; \emph{(source./dest.)}\emph{u.domid} stores the identification number of the SD/DD; \emph{len} stores the number of bytes to be copied; \emph{flags} stores a value indicating whether the SD and the DD are local or remote domains.

\smallskip
\textbf{Output:}\textsuperscript{\ref{fnt:20131964}}  On success, \emph{GNTTABOP\_copy} returns 0. On failure, \emph{GNTTABOP\_copy} returns an error code (typically a negative integer value). \emph{(struct gnttab\_copy)}\emph{status} stores a value indicating the status of the page copying operation.

\smallskip
\textbf{Workflow of the vulnerable hypercall handler:}\textsuperscript{\ref{fnt:20131964}}

\begin{lstlisting}[breaklines=true, backgroundcolor=\color{light-gray},escapeinside=`', basicstyle=\small, frame=tb,columns=fullflexible]
i `$\leftarrow$' the domain invoking GNTTABOP_copy
d `$\leftarrow$' the DD

do_grant_table_op(GNTTABOP_copy, struct grant_table_op op, ...)
  `\textbf{call}' gnttab_copy(op, ...)
    `\textbf{call}' __gnttab_copy(op, ...)
      . . .
      `\textbf{if}' the DD is remote
         `\textbf{call}' __acquire_grant_for_copy
          . . .
          act = active grant table entry (op.dest.ref)
          . . .
          `\textbf{if}' the grant to be acquired is non-transitive
            . . .
            `\textit{\textbf{act.trans\_domain = i}}' 
            `\textit{\textbf{act.trans\_gref = 0}}'
          . . .
         `\textbf{return}'
      . . .
      `\textbf{if}' the DD is remote
        `\textbf{call}' __release_grant_for_copy(d, op.dest.ref, ...)
          .  .  .
          act = active grant table entry (op.dest.ref)
          .  .  .
          `\textbf{if}' the grant to be released is of version 2
            `\textit{\textbf{if act.trans\_domain != d}}'
              `\textit{\textbf{call \_\_release\_grant\_for\_copy(act.trans\_domid, act.trans\_gref, ...)}}'
        `\textbf{return}'
      . . .
    `\textbf{return}'
  `\textbf{return}'
`\textbf{return}'
\end{lstlisting}

\smallskip
\textbf{Description of the vulnerability:} In the handler of the hypercall operation \emph{GNTTABOP\_copy}, the function \emph{\_\_acquire\_grant\_for\_copy} is used for acquiring grants and \emph{\_\_release\_grant\_for\_copy(d, gref, ...)} for releasing grants, where \emph{d} is the domain that has issued the grant to be released and \emph{gref} is the reference of the grant to be released. In case a grant of version 2 is acquired, the hypervisor creates an active grant and sets the values of its fields \emph{trans\_domid} and \emph{trans\_gref} to the identification number of the domain from where GNTTABOP\_copy has been invoked and 0, respectively. The reason for the latter is to enable scenarios involving, as described in the source code of the handler of \emph{GNTTABOP\_copy}, ``grant being issued by one domain, sent to another one, and then transitively granted back to the original domain''.

The way in which the scenario mentioned above is supported causes non-transitive grants of version 2 to be released as if they were transitive grants (i.e., in a recursive manner). 
The culprit of this error is that when releasing a grant in the handler of \emph{GNTTABOP\_copy}, it is assumed that a transitive grant is a grant whose \emph{trans\_dom} field stores a domain identification number that is not equal to the identification number of the domain that has issued the grant being released. However, since the value of the field \emph{trans\_domid} of a non-transitive grant is set to the identification number of the domain from where GNTTABOP\_copy has been invoked when the grant has been acquired, the previously mentioned condition is also true for non-transitive grants of version 2. As a result, when a non-transitive (active) grant of version 2 is released in the handler of \emph{GNTTABOP\_copy}, at least one more grant release takes place, where the grant with a grant reference 0, issued by the domain from where GNTTABOP\_copy has been invoked, is released.

An attacker can trigger CVE-2013-1964 by invoking GNTTABOP\_copy such that, for example, a page is copied from a local SD to a remote DD, which has issued a non-transitive grant of version 2. 

\smallskip
\textbf{Vulnerability fix:} A patch fixing the vulnerability CVE-2013-1964 was released on 18 April 2013 and is available at~\cite{patch20131964}. The patch modifies \emph{\_\_acquire\_grant\_for\_copy} such that the value of \emph{trans\_domid} is set to the identification number of the domain that issued the grant that is acquired. Further, the value of \emph{trans\_gref} is set to the reference of the grant that is acquired. These modifications of \emph{\_\_acquire\_grant\_for\_copy} prevent the recursive release of non-transitive grant of version 2.
 
\smallskip
\textbf{Triggering CVE-2013-1964:} We triggered CVE-2013-1964 in the following environment:

\begin{quote}
\begin{compactitem}[$\circ$]
\item guest VM - OS: Ubuntu Precise (32 bit), kernel 3.8.0-29-generic; 
\item host VM - OS: Ubuntu Precise (32 bit), kernel 3.8.0-29-generic; 
\item hypervisor - Xen 4.1.2.
\end{compactitem}
\end{quote}

\begin{figure}[ht]
\centering
\includegraphics[scale=1.0]{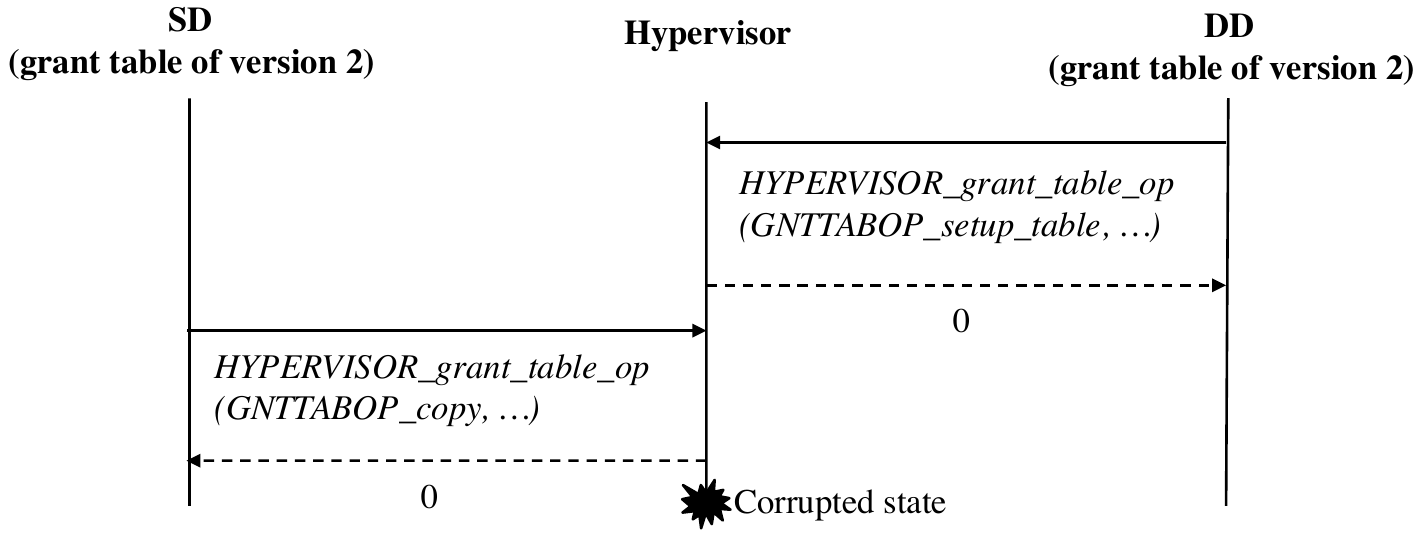}
\caption{An attack triggering CVE-2013-1964}
\label{fig:cve20131964}
\end{figure}

The attack that we executed is depicted in Figure~\ref{fig:cve20131964}.

\smallskip
\textbf{Post-attack state of the hypervisor:} Triggering CVE-2013-1964 results in a release of the grant with reference 0 issued by the domain from where GNTTABOP\_copy is invoked. Triggering CVE-2013-1964 may disrupt the operation of the hypervisor if the grant released due to the triggering of CVE-2013-1964 is in use (i.e., acquired) at the time of execution of the attack. When we triggered CVE-2013-1964 in our testbed environment, the hypervisor continued operating in a corrupted state.

\subsection{Hypercall set\_debugreg}

\paragraph{Vulnerability CVE-2012-3494} 

\begin{quote}
``The set\_debugreg hypercall in include/asm-x86/debugreg.h in Xen 4.0, 4.1, and 4.2, and Citrix XenServer 6.0.2 and earlier, when running on x86-64 systems, allows local OS guest users to cause a denial of service (host crash) by writing to the reserved bits of the DR7 debug control register.''~\cite{cve20123494}
\end{quote}

The \emph{set\_debugreg} hypercall is used for setting the value of the DR7 register of a CPU allocated to a guest VM. The DR7 register is used for controlling the actions of a CPU when program debugging is performed (e.g., for setting data and/or instruction breakpoints). The addresses at which breakpoints are set in a given debugging session are stored in the registers DR0 - DR3.

The layout of the DR7 register of a 64-bit machine is as follows: $^{bit63}$ 0 0 0 0 ....0 $^{bit31}$  [LEN3][R/W3] ... [LEN0][R/W0]$^{bit15}$ 0 0 $^{bit13}$ GD$^{bit11}$ 0 0 $^{bit9}$ GE LE $^{bit7}$ [G3][L3] ... [G0][L0]. The upper 32 bits are reserved and should always be cleared. The \emph{LENx} and \emph{R/Wx} fields are used for specifying the length of the monitored data items when a data breakpoint is set (e.g., 00: one-byte length - also when an instruction breakpoint is set, 01: two-byte length) and the type of program execution break set (e.g., 00 - instruction break, 01 - break on data write, 11 - break on data read and write), respectively. The \emph{GE} (global exact) and/or the \emph{LE} (local exact) bits are set when a data breakpoint is set and instruct the CPU to slow down the execution of the program being debugged so that the exact instruction that triggers the data breakpoint can be reported to the debugging program. The \emph{Gx} and \emph{Lx} bits are used for enabling or disabling breakpoints set at the addresses stored in the registers DR0 - DR3. 

\smallskip
\textbf{Input:}\footnote{As in Xen of version 4.1.2.\label{fnt:20123494}} \emph{set\_debugreg} takes as input a number of a register (an integer value, 7 is used for specifying the DR7 register) and a value that is to be stored in the register (an unsigned long integer value).

\smallskip
\textbf{Output:}\textsuperscript{\ref{fnt:20123494}} On success, \emph{set\_debugreg} returns 0. On failure, \emph{set\_debugreg} returns an error code (typically a negative integer value). 

\smallskip
\textbf{Workflow of the vulnerable hypercall handler:}\textsuperscript{\ref{fnt:20123494}} 

\begin{lstlisting}[breaklines=true, backgroundcolor=\color{light-gray},escapeinside=`', basicstyle=\small, frame=tb,columns=fullflexible]
do_set_debugreg (int reg_nr, unsigned long value)
  `\textbf{call}' set_debugreg(reg_nr, value)
    `\textbf{if}' reg_nr = 7   
       `\textit{\textbf{value}} \textbf{\&=} \textit{\textbf{\textasciitilde DR\_CONTROL\_RESERVED\_ZERO}}' 
       . . .
       `\textit{\textbf{store value in DR7}}'
  `\textbf{return}'
`\textbf{return}'
\end{lstlisting}

\smallskip
\textbf{Description of the vulnerability:} In the handler of the \emph{set\_debugreg} hypercall, the value of the variable \emph{\textasciitilde{}DR\_CONTROL\_RESERVED\_ZERO} is applied as a mask with the binary bitwise AND operator to the value of the second parameter of \emph{set\_debugreg}. The latter is performed so that the upper 32 bits of the value that is to be stored in the DR7 register are cleared. \emph{DR\_CONTROL\_RESERVED\_ZERO}, which stores the value of \emph{0x0000d800ul}, translates to the binary value of $^{bit63}$(0...0)$^{bit31}$(0000) (0000) (0000) (0000) (1101) (1000) (0000) (0000)$^{bit0}$. The complement form of the previously mentioned binary number is: $^{bit63}$(1...1)$^{bit31}$(1111) (1111) (1111) (1111) (0010) (0111) (1111) (1111)$^{bit0}$, which is stored in the variable \emph{\textasciitilde{}DR\_CONTROL\_RESERVED\_ZERO}. Since they are set to 1, the upper 32 bits of \emph{\textasciitilde{}DR\_CONTROL\_RESERVED\_ZERO} do not clear the upper 32 bits of the value that is to be stored in the DR7 register when applied as a mask with the binary bitwise AND operator. This results in setting one or multiple bits of the upper 32 bits of the DR7 register to 1, which is not allowed according to hardware specifications.  
  
CVE-2012-3494 can be triggered by invoking \emph{set\_debugreg} in a way such that one or multiple bits of the upper 32 bits of the value of the second parameter of \emph{set\_debugreg} are set to 1. The bits of the second parameter of \emph{set\_debugreg}   that are used for setting data or instruction breakpoints (e.g., the bits of the \emph{LENx} fields) should store binary values for setting an instruction breakpoint.

\smallskip
\textbf{Vulnerability fix:} A patch fixing the vulnerability CVE-2012-3494 was released on 5 September 2012 and is available at~\cite{patch20123494}. The patch assigns the value of \emph{\textasciitilde{}0xffff27fful} to \emph{DR\_CONTROL\_RESERVED\_ZERO}, and thus, the variable \emph{ \textasciitilde{}DR\_CONTROL\_RESERVED\_ZERO}, which is applied as a mask to the value of the second parameter of \emph{set\_debugreg}, has the binary value of $^{bit63}$(0...0)$^{bit31}$(0000) (0000) (0000) (0000) (0010) (0111) (1111) (1111)$^{bit0}$. Since the upper 32 bits of \emph{\textasciitilde{}DR\_CONTROL\_RESERVED\allowbreak\_ZERO} are cleared, applying \emph{\textasciitilde{}DR\_CONTROL\_RESERVED\allowbreak\_ZERO} as a mask to the value of the second parameter of \emph{set\_debugreg} with the binary bitwise AND operator clears the upper 32 bits of the parameter.

\smallskip
\textbf{Triggering CVE-2012-3494:} We triggered CVE-2012-3494 in the following environment:
\begin{quote}
\begin{compactitem}[$\circ$]
\item guest VM - OS: Ubuntu Precise (32 bit), kernel 3.8.0-29-generic; 
\item host VM - OS: Ubuntu Precise (32 bit), kernel 3.8.0-29-generic; 
\item hypervisor - Xen 4.1.2.
\end{compactitem}
\end{quote}

\begin{figure}[ht]
\centering
\includegraphics[scale=1.0]{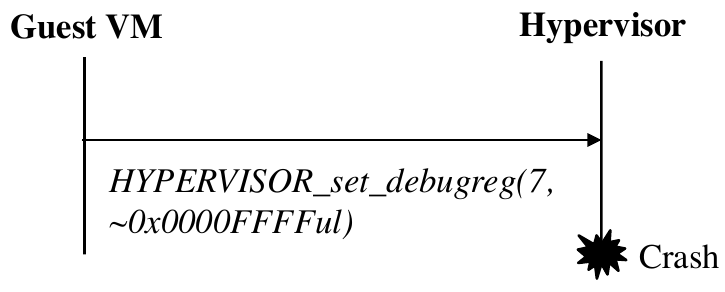}
\caption{An attack triggering CVE-2012-3494}
\label{fig:cve20123494}
\end{figure}

The attack that we executed is depicted in Figure~\ref{fig:cve20123494}.

\smallskip
\textbf{Post-attack state of the hypervisor:} Given that in current systems the upper 32 bits of the DR7 register are reserved and should be cleared, triggering CVE-2012-3494 results in crash of the vulnerable hypervisor. However, an outcome different than crash of the hypervisor may be possible if a vulnerable hypervisor is run on future hardware, as stated in \cite{cve20123496}: ``\emph{if the vulnerable hypervisor is run on future hardware, the impact of the vulnerability might be widened depending on the future assignment of the currently-reserved debug register bits.}''

\subsection{Hypercall physdev\_op}

\paragraph{Vulnerability CVE-2012-3495} 

\begin{quote}
``The physdev\_get\_free\_pirq hypercall in arch/x86/physdev.c in Xen 4.1.x and Citrix XenServer 6.0.2 and earlier uses the return value of the get\_free\_pirq function as an array index without checking that the return value indicates an error, which allows guest OS users to cause a denial of service (invalid memory write and host crash) and possibly gain privileges via unspecified vectors.''~\cite{cve20123495}
\end{quote}

\emph{PHYSDEVOP\_get\_free\_pirq} is an operation of the \emph{physdev\_op} hypercall, which is used for allocating PIRQ (PCI IRQs - Peripheral Component Interconnect Interrupt ReQuests) for the needs of a given guest VM. The Xen hypervisor maintains an array called \emph{pirq\_irq} for each guest VM that it hosts. \emph{pirq\_irq} is used for marking a given PIRQ as allocated such that the value of the constant variable PIRQ\_ALLOCATED (i.e., -1) is stored in the node of \emph{pirq\_irq} of index equal to the allocated PIRQ.

\smallskip
\textbf{Input:}\footnote{As in Xen of version 4.1.2.\label{fnt:20123495}}  \emph{PHYSDEVOP\_get\_free\_pirq} takes as input structure of type \emph{physdev\_get\_free\_pirq} defined as:

\begin{lstlisting}[breaklines=true, backgroundcolor=\color{light-gray},escapeinside=`', basicstyle=\small, frame=tb, columns=fullflexible, ]
struct physdev_get_free_pirq {
  int type;
  uint32_t pirq;
}
\end{lstlisting}

\emph{type} stores the type of the PIRQ to be allocated (i.e., \emph{MAP\_PIRQ\_TYPE\_GSI} or \emph{MAP\_PIRQ\_TYPE\_MSI}).

\smallskip
\textbf{Output:}\textsuperscript{\ref{fnt:20123495}} On success, \emph{PHYSDEVOP\_get\_free\_pirq} returns 0 and the allocated PIRQ is stored in \emph{(struct physdev\_get\_free\_pirq)} \emph{pirq}. On failure, -28 is stored in \emph{(struct physdev\_get\_free\_pirq)} \emph{pirq} and \emph{XENMEM\_populate\_physmap} returns an error code (typically a negative integer value). 

\smallskip
\textbf{Workflow of the vulnerable hypercall handler:}\textsuperscript{\ref{fnt:20123495}}
\begin{lstlisting}[breaklines=true, backgroundcolor=\color{light-gray},escapeinside=`', basicstyle=\small, frame=tb,columns=fullflexible]
do_physdev_op (PHYSDEVOP_get_free_pirq, (struct physdev_get_free_pirq) gfp)
  . . .
  `\textit{\textbf{call gfp.pirq = get\_free\_pirq(...)}}' 
    allocate a PIRQ
  `\textbf{return}'
  `\textit{\textbf{pirq\_irq[gfp.pirq] = PIRQ\_ALLOCATED}}' 
  . . .
`\textbf{return}'
\end{lstlisting}

\smallskip
\textbf{Description of the vulnerability:} In the handler of the \emph{PHYSDEVOP\_get\_free\_pirq} hypercall operation, the function \emph{get\_free\_pirq} is invoked for allocating a PIRQ. The return value of \emph{get\_free\_pirq} is the allocated PIRQ, if a PIRQ has been succesfully allocated, or an error code (i.e., -28) if a PIRQ could not be allocated. The return value of \emph{get\_free\_pirq} is used as an index to access an element of the array \emph{pirq\_irq} for marking a PIRQ as allocated. However, the return value of \emph{get\_free\_pirq} is not checked whether it is a PIRQ or an error code. In case \emph{get\_free\_pirq} returns an error code, the error code is used as an array index and as a result the value of the constant variable \emph{PIRQ\_ALLOCATED} (i.e., -1) is written at the memory address \&\emph{pirq\_irq} - 28, which is a location in hypervisor's memory.

CVE-2012-3495 can be triggered by attempting to allocate a PIRQ when there are no available PIRQs. This can be achieved by invoking the hypercall operation \emph{PHYSDEVOP\_get\_free\_pirq} multiple times until all available PIRQs are allocated and an attempt is made to allocate a PIRQ when there are no available PIRQs. Since PIRQs that can be allocated to a given VM are in the range of 16 to the value of the variable \emph{nr\_pirqs\_gsi}, a variable in hypervisor context that stores the largest PIRQ that can be allocated to a given VM, invoking \emph{PHYSDEVOP\_get\_free\_pirq} (\emph{nr\_pirqs\_gsi} - 16) + 2) times is sufficient for triggering CVE-2012-3495.

\smallskip
\textbf{Vulnerability fix:} A patch fixing the vulnerability CVE-2012-3495 was released on 5 September 2012 and is available at~\cite{patch20123495}. The patch inserts an \emph{if} clause that checks whether the return value of the \emph{get\_free\_pirq()} function is a PIRQ. If \emph{get\_free\_pirq()} returns an error code, the error code is not used as an index for accessing an element of the \emph{pirq\_irq} array. 

\smallskip
\textbf{Triggering CVE-2012-3495:} We triggered CVE-2012-3495 in the following environment:

\begin{quote}
\begin{compactitem}[$\circ$]
\item guest VM - OS: Ubuntu Precise (32 bit), kernel 3.8.0-29-generic; 
\item host VM - OS: Ubuntu Precise (32 bit), kernel 3.8.0-29-generic; 
\item hypervisor - Xen 4.1.2.
\end{compactitem}
\end{quote}

\begin{figure}[ht]
\centering
\includegraphics[scale=1.0]{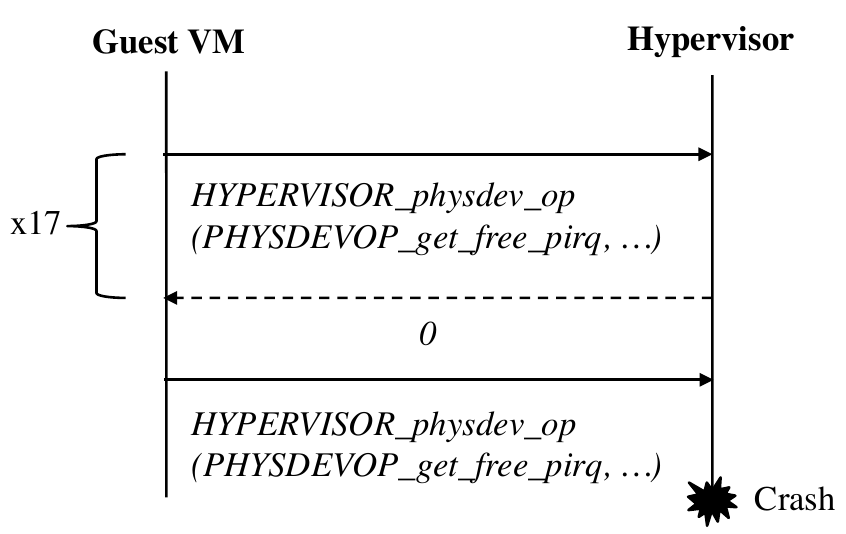}
\caption{An attack triggering CVE-2012-3495}
\label{fig:cve20123495}
\end{figure}

The attack that we executed is depicted in Figure~\ref{fig:cve20123495}.

\smallskip
\textbf{Post-attack state of the hypervisor:} Triggering CVE-2012-3495 results in overwriting the hypervisor's memory at the memory address \&pirq\_irq - 28 with the value of the variable PIRQ\_ALLOCATED (i.e., -1). An attacker cannot control the value written in the hypervisor's memory. Depending on the memory layout of the hypervisor, the hypervisor may crash or continue operating in a corrupted state. When we triggered CVE-2012-3495 in our testbed environment, the hypervisor continued operating in a corrupted state.   

\subsection{Hypercall mmuext\_op}

\paragraph{Vulnerability CVE-2012-5525} 

\begin{quote}
``The \emph{get\_page\_from\_gfn} hypercall function in Xen 4.2 allows local PV guest OS administrators to cause a denial of service (crash) via a crafted GFN that triggers a buffer over-read.''~\cite{cve20125525}
\end{quote}

The \emph{get\_page\_from\_gfn} function provides information about a given memory page. It is invoked in the handlers of multiple hypercalls of the Xen hypervisor, one of which is the handler of the \emph{MMUEXT\_CLEAR\_PAGE} operation of the \emph{mmuext\_op} hypercall. \emph{MMUEXT\_CLEAR\_PAGE} is an operation of the \emph{mmuext\_op} hypercall, which is used for clearing memory pages/frames. 

\smallskip
\textbf{Input:}\footnote{As in Xen of version 4.2.0.\label{fnt:20125525}} \emph{MMUEXT\_CLEAR\_PAGE} takes as input structure of type \emph{mmuext\_op} defined as:

\begin{lstlisting}[breaklines=true, backgroundcolor=\color{light-gray},escapeinside=`', basicstyle=\small, frame=tb, columns=fullflexible, ]
struct mmuext_op {
  unsigned int cmd;
  union {
  xen_pfn_t mfn;
  . . .
  } arg1;
  . . .
}
\end{lstlisting}

\emph{cmd} stores a number identifying an operation of the \emph{mmuext\_op} hypercall (e.g., \emph{MMUEXT\_CLEAR\_PAGE}); \emph{arg1.mfn} stores the MFN of the page that is to be cleared. 

\smallskip
\textbf{Output:}\textsuperscript{\ref{fnt:20125525}} On success, \emph{MMUEXT\_CLEAR\_PAGE} returns 0. On failure, \emph{MMUEXT\_CLEAR\_PAGE} returns an error code (typically a negative integer value).

\smallskip
\textbf{Workflow of the vulnerable hypercall handler:}\textsuperscript{\ref{fnt:20125525}} 

\begin{lstlisting}[breaklines=true, backgroundcolor=\color{light-gray},escapeinside=`', basicstyle=\small, frame=tb,columns=fullflexible]
do_mmuext_op ((struct mmuext_op) op, ...)
  struct page_info page;
  `\textit{\textbf{call page = get\_page\_from\_gfn(op.arg1.mfn, ...)}}'
  . . .
`\textbf{return}'
\end{lstlisting}

\smallskip
\textbf{Description of the vulnerability:} \emph{get\_page\_from\_gfn} reads information about a page allocated to a guest VM from the frame table of the VM using the MFN of the page as offset. A frame table is a memory area shared between the hypervisor and a guest VM where information about each page allocated to the guest VM is stored in the format of a structure of type \emph{page\_info}. 

The MFN used by \emph{get\_page\_from\_gfn} for reading page information is provided to \emph{get\_page\_from\_gfn} as an input parameter. The value of the MFN provided as input parameter to \emph{get\_page\_from\_gfn} is not checked for validity. Since \emph{get\_page\_from\_gfn} uses a MFN as an offset for reading from the frame table of a given guest VM, an invalid MFN is a MFN that causes a buffer over-read (i.e., that is larger than the largest MFN at which a page of the guest VM is allocated). An attacker can provide an invalid MFN as an input parameter to \emph{get\_page\_from\_gfn}, in which case \emph{get\_page\_from\_gfn} returns invalid page information. 

In the handler of the \emph{MMUEXT\_CLEAR\_PAGE} hypercall operation, the MFN stored in the input parameter \emph{(struct mmuext\_op) arg1.mfn} is provided to \emph{get\_page\_from\_gfn} for reading page information. CVE-2012-5525 can be triggered by invoking \emph{MMUEXT\_CLEAR\_PAGE} such that an invalid MFN is stored in \emph{(struct mmuext\_op) arg1.mfn}. 

\smallskip
\textbf{Vulnerability fix:} A patch fixing the vulnerability CVE-2012-5525 was released on 3 December 2012 and is available at~\cite{patch20125525}. The patch inserts an invocation of the function \emph{mfn\_valid} in \emph{get\_page\_from\_gfn}, which verifies the validity of the MFN provided as input to \emph{get\_page\_from\_gfn}. The patch modifies \emph{get\_page\_from\_gfn} such that if the MFN used for reading page information is not valid, \emph{get\_page\_from\_gfn} returns NULL instead of invalid page information. 

\smallskip
\textbf{Triggering CVE-2012-5525:} We triggered CVE-2012-5525 in the following environment:

\begin{quote}
\begin{compactitem}[$\circ$]
\item guest VM - OS: Ubuntu Precise (32 bit), kernel 3.8.0-29-generic; 
\item host VM - OS: Ubuntu Precise (32 bit), kernel 3.8.0-29-generic; 
\item hypervisor - Xen 4.2.0.
\end{compactitem}
\end{quote}

\begin{figure}[ht]
\centering
\includegraphics[scale=1.0]{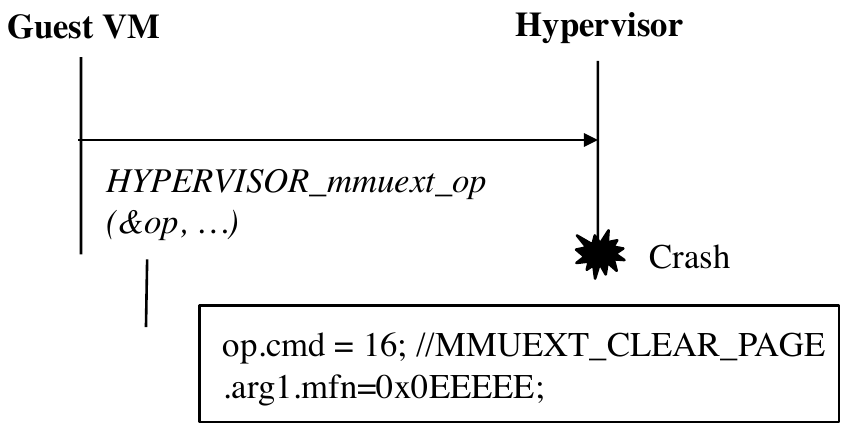}
\caption{An attack triggering CVE-2012-5525}
\label{fig:cve20125525}
\end{figure}

The attack that we executed is depicted in Figure~\ref{fig:cve20125525}.

\smallskip
\textbf{Post-attack state of the hypervisor:} Triggering CVE-2012-5525 may result in a crash of the targeted hypervisor or may corrupt its state. Whether the hypervisor crashes depends on the use of the invalid page information returned from \emph{get\_page\_from\_gfn} when CVE-2012-5525 is triggered. The hypervisor crashed when we triggered CVE-2012-5525 in our testbed environment.

\renewcommand\bibname{References}
\addcontentsline{toc}{section}{\bibname}
\bibliographystyle{plain}
\bibliography{SPEC-RG-2014-001_HypercallVulnerabilities}

\end{document}